
\magnification=\magstep1
\voffset=0.0truemm
\hoffset=0.0truemm
\hsize=6truein
\vsize=8.5truein
\raggedbottom
\baselineskip=24.0truept
\parindent=2.5em
\font\it=cmti12 at 12truept
\font\tenrm=cmr10 at 10truept
\font\tenit=cmti10 at 10truept
\centerline{\bf CYLINDRICAL BLACK HOLE IN GENERAL RELATIVITY}
\medskip
\centerline{\bf Jos\'e P. S. Lemos}
\centerline{\tenit  Departamento de Astrof\'{\i}sica, Observat\'orio
Nacional-CNPq, Rua General Jos\'e Cristino77,}
\centerline {\tenit 20921 Rio de Janeiro, Brazil,}
\centerline{\tenit \& Departamento de F\'{\i}sica, Instituto Superior
T\'ecnico, Av. Rovisco Pais 1, 1096 Lisboa, Portugal.}

\centerline{\tenrm ABSTRACT}
$$\vbox{\baselineskip=14truept\hsize 5truein\tenrm \noindent
A black hole solution of Einstein's field
equations with cylindrical symmetry is found. Using the Hamiltonian
formulation one is able to define mass and angular momentum for the
cylindrical black hole through the corresponding and equivalent
three dimensional theory. The causal structure is analyzed.}$$
\vskip -25.5truept

\bigskip\bigskip
\noindent{\bf 1. Introduction}

In a classical framework black hole solutions are of great
relevance since they might reflect the manner into which spacetime
settles after complete gravitational collapse of some form (e.g.,
collapse of stars or clusters of stars) has ocurred [1,2].
At the quantum level black holes are being used as theoretical
laboratories in the sense that they can give clues for the underlying
nature of the interaction between the geometry and the quantum world.
Examples of this interaction are the Hawking radiation [3],
scattering processes involving particles and black holes [4],
and the statistics a black hole gas should obey [5].

In General Relativity the black hole solutions which have so far been
found form a four parameter family called the generalized Kerr-Newman
family of black holes. The four parameters are mass $M$, angular momentum
$J$, charge $Q$, and the cosmological constant $\Lambda$ [6].
These are axisymmetric solutions and depending on the cosmological
constant have different asymptotic behavior, they can be asymptotically
flat ($\Lambda=0$), de Sitter ($\Lambda>0$) or anti-de Sitter
($\Lambda<0$).

Axial symmetry has two important particular cases. One is spherical
symmetry which has been extensively studied through the Schwarzschild
solution and the Schwarzschild black hole. The other is cylindrical
symmetry. Within the field of exact solutions, cylindrical symmetry
has played an important role in the discussion of the internal consistency
of General Relativity itself, through the static solutions of Levi-Civita
[7,8] and Chazy-Curzon [9,10], and the stationary solutions of
Lewis [11]. In an astrophysical context, cylindrical symmetry has been
applied to the study of cosmic strings [12] which in turn
offered a gain in understanding  the role conical singularities
and spacetime topological defects play in the gravitational field.
However, up to now there is no black hole solutions with cylindrical
symmetry. In this work we show that cylindrically symmetric rotating
black hole solutions of Einstein's field equations with a negative
cosmological constant do indeed exist.

\bigskip\bigskip
\noindent{\bf 2. The equations and the solution}

Einstein-Hilbert action in four dimensions is given by
$$ S={1\over 16\pi G}\int d^4x\sqrt{-g}\left( R-2\Lambda\right),
\eqno(1)$$
where $g$ is the determinant of the metric and $R$ the Ricci
scalar. We assume that the spacetime is cylindrically symmetric
and time-independent, i.e, there are three Killing vectors [13]:
$\partial\over{\partial z}$ which corresponds to the translational
symmetry along the axis, $\partial\over{\partial \varphi}$ which has
closed periodic trajectories around the axis and
$\partial\over{\partial t}$ corresponding to the invariance under
time translations.
We then find that the following solution
satisfies the equations of motion derived from (1),
$$ds^2 =
- \left( \alpha^2r^2 - {b\over{\alpha r}}\right) d\overline{t}^2
+{dr^2\over{\alpha^2r^2 - {b\over{\alpha r}}} }
+r^2d{\overline{\varphi}}^2
+\alpha^2 r^2 dz^2, \eqno(2)$$
$$-\infty< {\overline t}<\infty,\quad 0\leq r<\infty,
\quad 0\leq{\overline \varphi<2\pi},\quad -\infty< z<\infty.$$
Here $r$ is the radial circunferential coordinate,
$\alpha^2\equiv -{1\over3}\Lambda>0$ and
$b$ is a constant which
as we will see is related to the mass and we now assume positive.
Equation (2) represents a static black hole with an  event
horizon at $\alpha r=b^{1\over3}$. Since the Kretschmann scalar is given by
$R_{abcd}R^{abcd}= 24\alpha^4\left( 1 + {b^2\over{2\alpha^6 r^6}}\right)$
there is a scalar polynomial singularity
at $ar=0$ .
To add angular momentum to the spacetime we perform the following
coordinate transformation
$$\overline{t} = \lambda t -{\omega\over\alpha^2}\varphi,\eqno(3)$$
where $\omega$ and $\lambda$ are constant parameters.
In order to get rid of minor coordinate difficulties we still have to
change to rotating axes by doing,
$$\overline{\varphi}=\lambda\varphi-\omega t. \eqno(4)$$
Substituting (3) and (4) into (2) we obtain
$$ds^2 =
- \left\lbrack \left(\lambda^2-{\omega^2\over\alpha^2}\right)\alpha^2r^2 -
{b\lambda^2\over{\alpha r}}\right\rbrack dt^2
-{ {\omega b}\over {\alpha^3 r} } 2 d\varphi dt
+{dr^2\over{\alpha^2r^2 - {b\over{\alpha r}}} }+$$
$$\quad\quad\quad+ \left\lbrack \left(\lambda^2-
{\omega^2\over\alpha^2}\right)
r^2 +{ {\omega^2 b}\over {\alpha^5 r} }\right\rbrack
d\varphi^2
+\alpha^2 r^2 dz^2, \eqno(5)$$
$$-\infty< {t}<\infty,\quad 0\leq r<\infty,
\quad 0\leq{\varphi<2\pi},\quad -\infty< z<\infty.$$
This represents a stationary cylindrical
black hole and, of course, also solves (1), as can be checked directly
through the equations of motion generated by action (1).
One can then be tempted to say that by inverting the coordinate
transformations (3) and (4) one gets back the static spacetime (2). On a
first glance this is indeed the case. However, transformation (3) is not a
permitted gobal coordinate transformation. This is shown in a clear
way in the work of Stachel [14,15].
Transformation (3) can be done locally, but not globally.
Spacetime is not simply connected. This means that
the first Betti number of the manifold is one since
closed curves encircling
the horizon cannot be shrunk to zero. This happens in either spacetime,
static given by (2)
and stationary given by (5). Both are
homeomorphic to each other and homeomorphic to
$\lbrace R^3-R\rbrace {\rm{x}} R$. Now, in both spacetimes there is a
timelike Killing field $\xi={ {\partial\over\partial t}} $.
In the static spacetime this corresponds to an exact
one-form $\overline{V}$ inverse to $\xi$ (i.e., $\overline{V}_\mu\equiv
{\xi_\mu\over \mid \xi\mid^2}$) given then by $\overline{V} = dt$
(see [14] for details). In the stationary spacetime
the corresponding one form is $\overline{V} =
dt  + {\omega\over\alpha^2} d\varphi$ which is a closed one-form but
not exact. De Rham's
cohomology theorems then state that, since the first Betti number of
the manifolds is one, there are global diffeomorphisms which map the
$\xi$ of the two manifolds, but there is no such  global
diffeomorphisms mapping $\overline{V}$ and $V$. Since the metric
maps vectors into one-forms it means that metrics (2) and (5) cannot be
globally mapped into each other. In this case, the map is given by
eqution (3) which is immediatly understood as a local map. This is
because $\varphi$ is a periodic coordinate which in turn requires time
to be also periodic. Thus, metrics (2) and (5) can
be locally mapped into each other but not globally, and therefore they
are distinct. As suggested by Stachel
[14] this distinction could be tested by an Arahanov-Bohm
type experiment. We have used the coordinate transformation trick
(3) and (4) to convert (2) into (5). But once we realize it is a
trick we cannot go back. Spacetime (5) gives a stationary spacetime,
while (2) gives a static one.
Note that in the Schwarzschild solution closed curves
can always be shrunk to a point. So this type of coordinate transformation
will not generate a rotating black hole, but a rotating infinite
string superposed on a Schwarzschild black hole.

Linet [16] has found the general static solution of Einstein's field
equations with cylindrical symmetry and cosmological constant. However
Linet's solution uses a coordinate system which looses the black hole.
Santos [17] has generalized Linet's solution for stationary fields.
Since the coordinate system used is related to Linet's it also looses
the black hole.

\bigskip\bigskip
\noindent{\bf 3. Definition of mass and angular momentum}

We now tackle the delicate issue of the mass and angular momentum
of the black hole. Asymptotically, as $r\rightarrow\infty$, the black
hole spacetime is not Minkowski but anti-de Sitter. As shown by
Henneaux and Teitelboim [18] one can give meaningful definitions
for fields that approach at large spacelike distances the anti-de Sitter
configuration, whose group of motions is O(3,2). However there are two
problems here. Firstly, for large $\pm z$ (keeping $r$ fixed) the black hole
does not approach the anti-de Sitter solution. Secondly, in a spacetime
in which the singularity extends uniformly over the infinite $z$-line
one expects that the total energy (i.e., the ADM mass) is infinite. We
now show that one can deal with both difficulties simultaneously. Since the
trouble lies in the infinity of the $z$ direction we have to find a
procedure to eliminate the $z$ coordinate altogether. To this end we
reccur to
an equivalent three-dimensional (3D) theory, i.e., a 3D theory which
reproduces the equations of motion of cylindrically symmetric General
Relativity.

The most general metric with one Killing vector, invariant under
$z\rightarrow -z$ can be written as [13],
$$ds^2=g_{ab}dx^adx^b + e^{-4\phi}dz^2, \eqno(6)$$
Where $a,b=t,r,\varphi$, $g_{ab}=g_{ab}\left( t,r\right)$ is the 3D
metric and
$\phi=\phi\left( t,r\right)$.  In fact the most general metric
includes another metric
function $A$ in which case the metric is not invariant
under $z\rightarrow-z$, see ref. [13]. For our purposes it is enough to put
$A=0$.

 From standard dimensional reduction
techniques on (6) and (1) we obtain the following 3D action,
$$S={1\over2\pi}\int d^3x \sqrt{-g} e^{-2\phi}\left( R-2\Lambda\right).
\eqno(7)$$
The $z$ dimension has disappeared but left its mark on the dilaton field
$\phi$. The field $A$ commented above would appear as a gauge field in
the dimensional reduction process. The theory with action (7) could
be called $\Omega=0$ 3D
Brans-Dicke theory (where $\Omega$ is the Brans-Dicke parameter) or
3D Teitelboim-Jackiw theory (since these authors proposed (7) for the
2D case [19,20]). By varying action (7) with respect to
$g_{ab}$ and $\phi$ one obtains equations of motion identical to
cylindrically symmetric General Relativity as given by action (1).
Of course the solutions of the equations of motion can be
transferred from one theory to the other. In, particular there is
the 3D black hole solution obtained directly from (5) which now reads
$$ds^2 =
- \left\lbrack \left(\lambda^2-{\omega^2\over\alpha^2}\right)\alpha^2r^2 -
{b\lambda^2\over{\alpha r}}\right\rbrack dt^2
-{ {\omega b}\over {\alpha^3 r} } 2 d\varphi dt
+{dr^2\over{\alpha^2r^2 - {b\over{\alpha r}}} }+$$
$$\quad\quad\quad+ \left\lbrack \left(\lambda^2-
{\omega^2\over\alpha^2}\right)
r^2 +{ {\omega^2 b}\over {\alpha^5 r} }\right\rbrack
d\varphi^2,
 \eqno(8)$$
$$e^{-2\phi}=c\alpha r, \eqno(9)$$
where $c$ is a dimensionless constant, $c>0$. The point is that
the black hole solution (8)-(9)
allows a meaningful definiton of (ADM) mass and angular momentum.
Spacetime is
asymptotically anti-de Sitter, now with O(2,2) as the group of motions.
The calculations needed to find the mass and angular momentum are
similar to those related to the 3D black hole of Ba\~nados, Henneaux,
Teitelboim and Zanelli [21]. We mention here that Horowitz and Welch
[22] also arrived at the stationary black  hole of 3D General
Relativity through a coordinate tranformation of the static black
hole.
With that black hole of 3D General Relativity one can apply almost
directly the formalism developed by Regge and Teitelboim [23,24].
Here, there is
an extra dilaton field.

In order to find the Hamiltonian formulation of the black hole
we write the metric in the canonical form
$$ds^2=-{N^0}^2dt^2 + R^2 \left( N^\varphi dt+d\varphi\right)^2
+{dR^2\over f^2}, \eqno(10)$$
where
$$R^2\equiv \left(\lambda^2- {\omega^2\over\alpha^2}\right) r^2
+ {\omega^2b\over\alpha^5 r},\quad\quad
{N^0}^2=\left( \alpha^2 r^2-  {b\over \alpha r}\right)
{\left(\lambda^2-{\omega^2\over\alpha^2}\right)^2}{ r^2\over R^2}
$$
$$
N^\varphi =  -{ \lambda\omega b\over\alpha^3 R^2 r} + {\rm constant},\quad
\quad
f^2=\left(\alpha^2 r^2 - {b\over\alpha r}\right) \left(
{dR\over dr}\right)^2.\eqno(11)$$
$N^0$ and $N^\varphi$ are the lapse and shift functions. By the usual
procedure [21,23] one can bring action (7) with the help of (8)-(10)
into the Hamiltonian form, which reads
$$S= -\Delta t \int \Bigl\{
N\left\lbrack
{1\over2} {e^{-2\phi} R^3 \left({N^\varphi}_{,R}\right)^2\over N^2 }
+ e^{-2\phi} \left( f^2\right)_{,R} \left(1-2R{d\phi\over dR}\right)-\right.$$
$$\left.-4f^2\left( e^{-2\phi} R {d\phi\over d R}\right)_{,R} +
2 e^{-2\phi}R\Lambda \right\rbrack+
N^\varphi\left\lbrack e^{-2\phi} R^3 {\left( {N^\varphi}_{,R}\right)\over N}
\right\rbrack_{,R} \Bigr\} dR + B.\eqno(12)$$
$N\equiv{N^0\over f}$ and $N^\varphi$ are Lagrange
multipliers imposing constraints on the action, namely the terms
inside the squared brackets should be zero.
$B$ is a surface term needed to ensure that Hamilton's equations are
satisfied. Our task is to find $B$ and associate it with the mass and
angular momentum.  To obtain the equations
of motion one has to vary (12) with respect to $\phi$, $f^2$ and the
momentum conjugate to the metric
$\pi=R^3{\left( N^\varphi\right)_{,R}\over N}$. Here, we
are interested only in the surface terms that one acquires by the variation
of the action [25]. That is, as $R\rightarrow\infty$, one finds
$$\delta S=
-\Delta t \left\lbrack N\left(\infty\right) e^{-2\phi}
\left(1-2R{d\phi\over d R}\right)\delta f^2\right\rbrack_{R=\infty}+$$
$$\quad\quad\quad -\Delta t \left\lbrack -4  N\left(\infty\right) f^2
\delta \left( e^{-2\phi} R
{d\phi\over d R}\right)\right\rbrack_{R=\infty}+$$
$$\quad\quad\quad -\Delta t  \left\lbrack 2 R e^{-2\phi}
\left( N\left(\infty\right) {df^2\over dR} -
{N^\varphi R^2 {N^\varphi}_{,R}\over N}\right)
\delta\phi
\right\rbrack_{R=\infty}+$$
$$\quad\quad\quad -\Delta t \left\lbrack N^\varphi\left(\infty\right)
e^{-2\phi}
\delta\left( {R^3 {N^\varphi}_{,R}\over N}\right)\right\rbrack_{R=\infty}
+\delta B.\eqno(13)$$
We have $\delta f^2= {f^2}_{BH} - {f}^2_{AdS}$, etc, (BH=black hole, AdS
=anti-de Sitter). Then by carefully examining each term in (13) we
arrive at
$$\delta S =
\Delta t \left\lbrack N\left(\infty\right) \delta \left(
bc\left(2\lambda^2+{\omega^2\over\alpha^2}\right)\right)
-N^\phi\left(\infty\right) \delta\left( 3bc\lambda{\omega\over\alpha^2}
\right)\right\rbrack + \delta B.\eqno(14)$$
Thus $\delta B$ has to be equal to minus the first term on the right hand
side of (14). So the boundary term $B$ is well defined and Hamilton's
equations follow. Since  mass and angular momentum are defined as
the terms
conjugated to $N$ and $N^\phi$ we have,
$$M= bc\left(2\lambda^2 + {\omega^2\over\alpha^2}\right),\eqno(15)$$
$$J=3bc\lambda{\omega\over\alpha^2}.\eqno(16)$$
Note that $M$ and $J$ depend on the strength of the dilaton through
the constant $c$, as it happens with 2D dilaton gravity theories [26,27,28].
Solving for $\lambda$ and $\omega\over\alpha$ yields
$$\lambda^2 = {1\over b} {M+M\sqrt{1-{8J^2\alpha^2\over9M^2}}\over 4c},
\eqno(17)$$
$${\omega^2\over\alpha^2}={1\over b} {M-M\sqrt{1-{8J^2\alpha^2\over9M^2}}
\over 2c}.\eqno(18)$$
In (17) we have taken the + sign in front of the descriminant (equation
(18) has the the corresponding $-$ sign).
Now, choosing $b$ is choosing a scale for the coordinate $r$. In order to
have the standard form of the anti-de Sitter spacetime at spatial
infinity we have to set
$b={1\over4c}\left(-M+3M\sqrt{1-{8J^2\alpha^2\over9M^2}}\right)$. Then the
3D black hole (8)-(9) is
$$ds^2=-\left( \alpha^2 r^2 -
{M+M\sqrt{1-{8J^2\alpha^2\over9M^2}}\over 4c\alpha r}\right) dt^2
-{J\over3\alpha r} 2dtd\varphi+$$
$$\quad\quad\quad+{ dr^2\over
\alpha^2 r^2 -
{-M+3M\sqrt{1-{8J^2\alpha^2\over9M^2}}\over4c\alpha r}  } +
\left( r^2 +
{M-M\sqrt{1-{8J^2\alpha^2\over9M^2}}\over2c\alpha^3 r}\right) d\varphi^2,
\eqno(19)$$
$$e^{-2\phi}=c \alpha r.\eqno(20)$$
By a coordinate transformation one can put $c=1$.
The cylindrical black hole
solution is then given by,
$$ds^2=-\left( \alpha^2 r^2 -
{M+M\sqrt{1-{8J^2\alpha^2\over9M^2}}\over 4\alpha r}\right) dt^2
-{J\over3\alpha r} 2dtd\varphi+$$
$$\quad\quad\quad+{ dr^2\over
\alpha^2 r^2 -
{-M+3M\sqrt{1-{8J^2\alpha^2\over9M^2}}\over4\alpha r}  } +
\left( r^2 +
{M-M\sqrt{1-{8J^2\alpha^2\over9M^2}}\over2\alpha^3 r}\right) d\varphi^2+$$
$$\quad\quad\quad+\alpha^2 r^2 dz^2.
\eqno(21)$$
The mass-energy of this system is  a mass-energy per unit length, which
is the meaningful quantity in cylindrical systems.  According to the
literature it can be considered more appropriate to call (21) a
(straight) black string instead of a black hole, although it seems to
us that in 4D the names black string and cylindrical black hole are
synonymous.
\vfill\eject
\bigskip\bigskip
\noindent{\bf 4. Causal Structure}

Now we turn to the causal struture which has many interesting aspects.
First assume that $r\geq0$. Then there are three
distinct cases with real solutions.
(i) $0\leq J\alpha<M$: this is the black hole solution.
There is an event horizon located at
${r_{eh}}^3= {1\over\alpha^3}
{-M+3M\sqrt{1-{8J^2\alpha^2\over9M^2}}\over 4}$. The infinite
redshift surface, always outside the horizon, is at
${r_{rs}}^3= {1\over\alpha^3}
{M+M\sqrt{1-{8J^2\alpha^2\over9M^2}}\over 4}$. There is another
important radius which gives the place at which upon decreasing $r$
the perimeter starts to increase. This turning point is at
${r_{tp}}^3= {1\over\alpha^3}
{M-M\sqrt{1-{8J^2\alpha^2\over9M^2}}\over 4}$. When
$0\leq J\alpha
\leq {3\sqrt3\over4\sqrt2}M$ then $r_{tp}\leq r_{eh}$.
For ${3\sqrt3\over4\sqrt2}M
< J\alpha < M$ one has $r_{eh}< r_{tp}$. At $r=0$ there is a spacelike
singularity. For $r\rightarrow\infty$ spacetime is anti-de Sitter.
(ii) $J\alpha=M$: there is a null singularity at $r =0$. This is
the extremal limit of the black hole.
(iii) $M<J\alpha\leq{\sqrt{9\over8}}M$: there are no horizons.
The singularity is timelike and naked.
Like Kerr's the singularity has a ring like structure. Unlike Kerr's
one cannot penetrate through the inside of the ring since the black hole
is 3D (and in 4D the symmetry is cylindrical).
There is an infinite redshift surface.
The corresponding Penrose diagrams are very simple and we do not
draw them here. For $J\alpha>{\sqrt{9\over8}}M$ the solution turns
complex.

There is another set of solutions when we do $r\rightarrow-r$ in (14)
or (16). In this set there are also three distinct cases all of them
have closed timelike curves.  Therefore, if one wants
chronology protection these cases should be discarded.

\bigskip\bigskip
\noindent{\bf 5. Temperature}

To include quantum field effects on the classical geometry one must
compute the temperature, the entropy and other associated potentials.
To display beyond doubt what corresponds to the extreme black hole
we find here the temperature $T$ as a function of $M$ and $J$.
By Euclideanizing metric (21) one can show that
$$
T={\alpha\over2\pi} {3\over2}
\left(   {3M\sqrt{1 -  {8J^2\alpha^2\over9M^2}}-M\over4}\right)^{1\over3}
\left({  3\sqrt{1 - {8J^2\alpha^2\over9M^2}}-1\over
\sqrt{1-  {8J^2\alpha^2\over9M^2}} +1 }\right)^{1\over2}.\eqno(22)$$
For $J=0$ the temperature goes with $M^{1\over3}$. Thus it tends to zero
as the horizon disappears. This is analogous to the black hole of
3D General Relativity [29] and in contrast to the Schwarzschild black hole.
The extreme case is then $J\alpha=M$, for which $T=0$.
\bigskip\bigskip
\noindent{\bf 6. Conclusions}

The cylindrical black hole has a rich structure which can be further
explored at the quantum and the classical level. The extension to include
eletromagnetic fields is under study [25]. It is remarkable
that the 3D version has given us insights into ill-defined quantities
(such as mass and angular momentum)
in the 4D spacetime.

\bigskip\bigskip
\vfill\eject
\bigskip

\centerline{References}
\item{1.} J. A. Wheeler, in {\it Relativity, Groups and Topology}, eds.
C. deWitt, B. S. deWitt (Gordon and Breach, 1964), p. 317.
\item{2.}
R. Penrose, {\it Phys. Rev. Lett.}, {\bf 14}, 57 (1965).
\item{3.} S. W. Hawking, {\it Nature}, {\bf 248}, 30 (1974).
\item{4.} J. Preskill, P. Schwarz, A. Shapere, S. Trivedi, F.
Wilczeck, {\it Mod. Phys. Lett. A}, {\bf 6}, 2353 (1991).
\item{5.} A. Strominger, {\it Phys. Rev. Lett.}, {\bf 71}, 3397 (1993).
\item{6.} B. Carter, in {\it Black Holes}, eds. C. DeWitt, B. S. DeWitt
(Gordon and Breach, 1973), p. 57.
\item{7.} T. Levi-Civita, {\it Rend. Acc. Lincei}, {\bf 26}, 317
(1917).
\item{8.} H. Weyl, {\it Annalen Phys.}, {\bf 54}, 117 (1917).
\item{9.} J. Chazy, {\it Bull. Soc. Math. France}, {\bf 52}, 17 (1924).
\item{10.}
H. E. J. Curzon, {\it Procc. R. Soc. London}, {\bf 23}, 477 (1924).
\item{11.} T. Lewis, {\it Proc. R. Soc. London}, {\bf A 136}, 176 (1932).
\item{12.} A. Vilenkin, {\it Phys. Rep}, {\bf 121}, 263 (1985).
\item{13.} D. Kramer, H. Stephani, M. MacCallum and E. Herlt,
{\it Exact Solutions of Einstein's Field Equations},
(Cambridge University Press, 1980).
\item{14.} J. Stachel, {\it Phys. Rev. D}, {\bf 26}, 1281 (1982).
\item{15.}
W. B. Bonnor, {\it J. Phys. A}, {\bf 13}, 2121 (1980).
\item{16.} B. Linet, {\it J. Math. Phys.}, {\bf 27}, 1817 (1986).
\item{17.} N. O. Santos,
{\it Class. Quantum Grav.}, {\bf 10}, 2401 (1993).
\item{18.} M. Henneaux, C. Teitelboim, {\it Comm. Math. Phys.}, {\bf 98},
391 (1985).
\item{19.} C. Teitelboim, in {\it Quantum Theory of Gravity, Essays in Honour
of the 60th Birthday of B. DeWitt}, ed. S. Christensen (Adam Hilger, Bristol,
1984), p. 327.
\item{20.} R. Jackiw,  in {\it Quantum Theory of Gravity, Essays in Honour
of the 60th Birthday of B. DeWitt}, ed. S. Christensen (Adam Hilger, Bristol,
1984), p. 403.
\item{21.} M. Ba\~nados, M. Henneaux, C. Teitelboim, J. Zanelli,
{\it Phys. Rev. D}, {\bf 48}, 1506 (1993).
\item{22.}
G. T. Horowitz, D. L. Welch, {\it Phys. Rev. Lett},
{\bf 71}, 328 (1993))
\item{23.} T. Regge, C. Teitelboim, {\it Ann. Phys.} (NY), {\bf 88}, 286
(1974).
\item{24.} A. Hanson, T. Regge, C. Teitelboim, {\it Constrained Hamiltonian
Systems}, Accademia Nazionale dei Lincei, Roma (1976).
\item{25.} J. P. S. Lemos, V. Zanchin, ``Charged Black String in General
Relativity''.
\item{26.} J. P. S. Lemos, P. M. S\'a, {\it Phys Rev. D}, {\bf 49}, 2897
(1994).
\item{27.} E. Witten, {\it Phys. Rev. D}, {\bf 44}, 314 (1991).
\item{28.}
A. Bilal, I. I. Kogan, {\it Phys. Rev. D}, {\bf 47}, 5408 (1993).
\item{29.}  M. Ba\~nados, C. Teitelboim, J. Zanelli,
{\it Phys. Rev. Lett}, {\bf 69}, 1849 (1993).

\end